\documentclass[prd,preprint,a4paper,superscriptaddress,nofootinbib,11pt]{revtex4}
\usepackage{amsmath,amsfonts,amssymb}
\usepackage{txfonts}
\usepackage[T1]{fontenc}
\usepackage[safe]{textcomp}

\usepackage{csquotes}
\newtheorem{theorem}{Theorem}[section]

\begin{document}

\begin{center}

{\bf  \Large Fradkin-Bacry-Ruegg-Souriau vector in kappa-deformed space-time}
 
 \bigskip

Partha Guha
{\footnote{e-mail:partha@bose.res.in}}, \\  
S.N. Bose National Centre for Basic Sciences
JD Block, Sector III, Salt Lake
Kolkata - 700098, India
 \\[3mm]
E. Harikumar  {\footnote{e-mail: harisp@uohyd.ernet.in }} and Zuhair N. S. {\footnote{e-mail:zuhairns@gmail.com}}\\
School of Physics, University of Hyderabad,
Central University P O, Hyderabad-500046,
India \\[3mm] 

\end{center}
\setcounter{page}{1}
\bigskip

{We study presence of an additional symmetry of a generic central potential in the $\kappa$-space-time. An explicit construction of Fradkin and Bacry, Ruegg, Souriau (FBRS) for a central potential is carried out and the piece-wise conserved nature of the vector is established. We also extend the study to Kepler systems with a drag term, particularly Gorringe-Leach equation is generalized to the $\kappa$-deformed space. The possibility of mapping Gorringe-Leach equation to an equation with out drag term is exploited in associating a similar conserved vector to system with a drag term. An extension of duality between two class of central potential is introduced in the $\kappa$-deformed space and is used to investigate the duality existing between two class of Gorringe-Leach equations. All the results obtained can be retraced to the correct commutative limit as we let $a \rightarrow 0$.}

\bigskip

 
\bigskip


\newpage
 
 \section{Introduction}
Investigation of symmetries plays a pivotal role in our understanding of physical laws as they are intimately related to the existence of integrals of motion. One of the familiar examples is the rotational symmetry enjoyed by central potentials. Noether's theorem shows that the rotational symmetry implies the existence of a constant of motion, namely angular momentum. Historically, it have been noticed that the Kepler system possess an additional conserved quantity, namely, Laplace-Runge-Lenz vector, which arise from the geometrical nature of the orbit rather than from a Noether symmetry. Interestingly, a similar conserved quantity, Fradkin-Hill tensor is known to exist for the isotropic oscillator \cite{hill, frad1}.
 
It was shown that all dynamical systems in three dimensions are invariant under $SU(3)$ and $O(4)$ algebra \cite{muk}. An investigation of these potentials showed that they contain a larger symmetry group than the rotational group. To be specific, the Kepler problem and isotropic oscillator problem respect $O(4)$ and $SU(3)$ symmetry, respectively \cite{BRS}. A generalization of above result showing that all central potentials (in classical systems) enjoy this extended symmetry was obtained in \cite{frad2} and this symmetry has been ascribed to existence of constant  plane of orbit \cite{frad2}. This lead to several investigations trying to construct conserved quantities, more general than Laplace-Runge-Lenz vector. It was shown that every three dimensional dynamical systems involving central potentials do admit a conserved vector and this general vector has been constructed and analyzed in\cite{BRS,frad2}. Shortly afterwards, it was shown that this generalized conserved vector is multi-valued\cite{vbs, lhb}.
 Bacry, Ruegg and Souriau showed that such a vector is exceptionally one-valued, in the Kepler case, and corresponds generally to a piecewise conserved quantity, the Fradkin-Bacry-Ruegg-Souriau (FBRS) perihelion vector \cite{HM, grand1}. An extension of similar construction on curved manifolds was taken up in \cite{bell}.

In \cite{peres}, Peres rederived this constant vector using an approach different from the ones adopted in\cite{BRS,frad2}. The starting point of \cite{peres} was the requirement that a generalization of of Laplace-Runge-Lenz vector with arbitrary coefficients to be an integral of motion. This condition, imposed restrictions on these coefficients which are functions of the radial coordinate. Thus, these coefficients were shown to satisfy a set of coupled differential equations. Thus the problem of obtaining a conserved vector was reduced to finding solutions to these differential equations. Further, by analyzing these differential equations, it was shown that the conserved vector is multi-valued. This construction was re-visited in \cite{grand1} for the central potentials in 2-dimensional space, using complex coordinates. The correspondence between Fradkin's and Peres approaches have been studied by Yoshida \cite{Yoshida}. In this paper, we generalize the approach of\cite{grand1} to analyze the central 
potentials in the $\kappa$-deformed space-time, which is an example of a non-commutative space-time.

\smallskip

It was shown by Grandati et al. \cite{grand2} that every generalized Gorringe-Leach equation admits an associated FBRS vector which is globally conserved for the equations of a particular class. For the dualizable generalized Gorringe-Leach equations, Grandati et al. \cite{grand2} showed the image sets of the discontinuous FBRS vectors for two classes of power potential problems are dual images of each other. In this paper we study the non-commutative generalization of the Gorringe-Leach equations in the $\kappa$-space time also.

\bigskip

Non-commutative space-times and various models on such spaces are being studied in recent times \cite{pres1, pres2, scnc1, scnc2}. One of the reasons of this interest is the important role of non-commutative space-time in the context of quantum gravity \cite{qs}. $\kappa$-space-time\cite{luk} is an example of a non-commutative space-time whose co-ordinates satisfy the conditions 
\begin{equation}
[\hat{x}_0, \hat{x}_i]=a\hat{x}_{i} {\rm ~~and~~} [\hat{x}_i, \hat{x}_j]=0.
\end{equation}
which naturally appears in the discussions of deformed special relativity\cite{dsr}. Various aspects of this space-time has been studied in recent times and one of the approaches used was to map the models defined on the $\kappa$-deformed space-time to usual commutative space-time\cite{mel}. Recently, there have been many interesting works trying to understand the dynamics of central potential, both at classical and quantum level, in non-commutative space-time \cite{rom1, rom2, hari}.

\smallskip

In this paper, we investigate the additional symmetry of central potential problems in kappa-deformed space-time. 
FBRS vector for a generic central potential is constructed in the $\kappa$-deformed space. It is shown that the vector 
is piece-wise constant as in the case of commutative space and the necessary condition for its constancy is obtained. 
Kepler-type systems with drag term are also studied in the $\kappa$-space-time and the existence of an FBRS-like vector is 
shown. Arnold-Bohlin-Vasiliev duality, which establishes a correlation between two classes of power potentials and 
helps in generalizing the duality existing between Kepler potential and harmonic oscillator potential is generalised 
to the $\kappa$-space. The mapping between two classes of Gorringe-Leach equation, which exist in commutative case, 
is extended to the $\kappa$-space using the Arnold-Bohlin-Vasiliev duality.     

This paper is organized as follows. In the second section, we set up the platform for describing a generic central force problem in kappa-deformed space-time. We express the Hamiltonian for the the corresponding system in terms of commutative variables using a realization which connects the non-commutative variables with commutative variables \cite{hari2, hari3, hari4}. We obtain the radial motion for the generic central potential system, expressed in terms of complex coordinate system, and obtained an expression for velocity. In the third section, we used the results derived in the previouse section to set up $\kappa$-deformed FBRS vector in complex coordinate system. We then show that, in general, such a system admits a piece wise conserved quantity as in the case of commutative situation. The structure of the conserved quantity is obtained as a general case and it is shown that even in the presence of non-commutativity, the integrability of the system is intact. In the next section, we consider the 
Hamiltonian for reparametrised Gorringe-Leach equation. Here we generalizes the Gorringe-Leach equation to $\kappa$-space-time and shows that it also possess a piece-wise integral of motion. Section 5 establishes the Bohlin-Arnold-Vassiliev duality for the Gorringe-Leach equation with a power potential. We first summarise the results in the commutative space for a Gorringe-Leach equation with power potential and this was utilized to confirm the existence of duality between class of potentials (which generalizes Kepler potential and harmonic potential) in $\kappa$ space-time. Our concluding remarks are presented in section 6.

\section{Central force problems in $\kappa$-deformed space-time} 
In this section, we derive the Fradkin-Bacry-Ruegg-Souriau's (FBRS) vector for the central potentials in the $\kappa$-deformed space-time. By explicit construction this vector is guaranteed to be a conserved quantity. Our results are valid to all orders in the deformation parameter `$a$'.
 
$\kappa$-deformed space-time is an example for a Lie-algebraic type non-commutative space-time which naturally arise in the low energy limit of certain quantum gravity models \cite{qs}. We can re-express the coordintes of $\kappa$-deformed space-time in terms of commutative phase space variables \cite{hari2} as
\begin{equation}
\hat{x}^\mu = x^{\mu}+\alpha x^{\mu}(a.p)+\beta(a.x)p^{\mu}+\gamma a^{\mu}(x.p) \label{hatcor}.
\end{equation}
and corresponding momenta are given by
\begin{equation}
\hat{p}^\mu = p^{\mu}+(\alpha +\beta)(a.p)p^{\mu}+\gamma a^{\mu}(p.p) \label{hatmom}.
\end{equation}
Here, the real constants $\alpha$, $\beta$ and $\gamma$ satisfy the conditions
\begin{equation} 
\gamma-\alpha=1,~~
\alpha, \gamma, \beta \in \mathbb{R}. 
\end{equation}
 With the choice, $a^\mu = (a,\vec{0})$ and setting $\beta = 0$, we obtain ${\hat x}_i$ explicitly as
\begin{equation}
  \hat{x}^i = x^i + \alpha x^i a p^0.
\end{equation}
Using this, we find the (square of the) norm of the position vector in the $\kappa$-space-time as
\begin{equation}
  \hat{r}^2 = \hat{x}_{i}^{2} = r^2 (1 + 2a \alpha E_{0} + a^2 \alpha^2 E_{0}^{2}). \label{rad}
\end{equation}
where we have used the identification of $p_{0}$ with $E_{0}$. 

As in \cite{hpz}, we start with the $\kappa$-deformed Hamiltonian,
\begin{equation}
H = \frac{p^2}{2\tilde{m}} + U(\hat{r}) \label{hami}
\end{equation}
where the deformed mass is given by
\begin{equation}
\tilde{m} = \frac{m}{1+2\alpha m}.\label{deformedmass}
\end{equation}
Note that the form of Hamiltonian is derived by taking the non-relativistic limit of the $kappa$-deformed Casimir relation, $\hat{p}_\mu \hat{p}^\mu = m^2$ \cite{hpz}.
Using this Hamiltonian, we find the equations of motion to be 
\begin{eqnarray}
\tilde{m} \overset{..}{x_{i}} = \{ \dot{x_{i}}, H \} = -\frac{\partial U(\hat{r})}{\partial x_{i}} \label{eom}.
\end{eqnarray}
Using eqn.(\ref{rad}), we find the potential, to the first order in the non-commutative parameter as
\begin{equation}
U(\hat{r}) = U((1+\alpha a E^0)r)=U(r)+\alpha aE^0 r\frac{\partial E}{\partial r}=(1+\alpha a E^0 n)U(r) \label{pot}
\end{equation}
where we made the identification of $p_{0}$ with $E_{0}$ and $U(r)$ is assumed to be an $nth$ order polynomial in $r$.

From the equations (\ref{pot}) and (\ref{eom}), we find that the modification of the equation of motion due to kappa-deformation can be absorbed into an overall multiplicative coefficient which depends on the deformation parameter `$a$'. Except for this, the equation. remain exactly same as in the commutative space-time \cite{hpz}. It is also easy to see that in the limit $a \rightarrow 0$, we reproduce the result in the commutative space-time.

The construction of FBRS vector can be readily done by re-expressing the central force problem in the complex plane \cite{grand2}. We briefly summarise the essential steps of re-expressing the central potential problem in the complex plane. For this, we restrict ourselves to two-dimensional motion described by vector
\begin{eqnarray}
\vec{r} = (x(t), y(t))  
\end{eqnarray}
We introduce the complex coordinate as $z(t)=x(t)+iy(t)$. 
The potential $U(z,\bar{z})$ can be viewed as a real valued function on complex plane, $\mathbb{C}$. Note that,
\begin{eqnarray}
\nabla &\rightarrow &2 \frac{\partial }{\partial \bar{z}} \\
  \frac{\partial }{\partial \bar{z}} &=& \frac{1}{2} (\frac{\partial }{\partial x}+i\frac{\partial }{\partial y})
\end{eqnarray}
In terms of the complex coordinate, the equation of motion becomes
\begin{equation}
\tilde{m} \overset{..}{z} + 2 \frac{\partial U(z,\bar{z})}{\partial \bar{z}} =0 \label{eom1}
\end{equation}
It should be noted that the mass that appears in the above equation is given in eqn.(\ref{deformedmass}).
In the present case, the potentials $U(z,\bar{z})$ we consider, satisfy the condition
\begin{equation}
 U(z,\bar{z}) = U(\lvert z\rvert ) = U(r) 
\end{equation}
Thus it is easy to see that 
\begin{equation}
\frac{\partial U}{\partial \bar{z}} = \frac{\partial U}{\partial \bar{z}}   \frac{\partial U}{\partial r} =
\frac{z}{2r} \frac{\partial U}{\partial r}  
\end{equation}
Recalling that the angular momentum in $\kappa$-space-time is similar in form 
to that in commutative space-time, except for the modified mass factor\cite{hpz}, we write the angular momentum associated with the above central potential as
\begin{equation}
\vec{L} = \tilde{m} \vec{r} \times \dot{\vec{r}}  
\end{equation}
where ${\tilde m}$ is the deformed mass. In the complex notation, the above angular momentum has the form
\begin{equation}
L = \mathcal{I}m {\bar z}(t){\dot z}(t)) = \frac{\tilde{m}}{2i} (\bar{z}\dot{z}-\dot{\bar{z}}z)  \label{ang}
\end{equation}
Our major concern in the present paper is the dynamics of systems with a radial symmetry. Thus, it would be helpful to review the motion along radial direction. The radial equation we are interested is written in terms of the complex coordinates. For this, we note that with $r^2 = z \bar{z}$ and thus
\begin{eqnarray}
2r\dot{r} = \dot{z}\bar{z} + z\dot{\bar{z}} \\{\rm ~~and~~}
\dot{r} = \frac{1}{r} (z\dot{z}+i\frac{L}{\tilde{m}})  \label{r}
\end{eqnarray}
where eqn.(\ref{ang}) is used in arriving at the last equation above. Using
\begin{eqnarray}
z=r \exp{i\theta} \label{z} {~and~~} \exp{2i\theta} = \frac{z}{\bar{z}},   
\end{eqnarray}
we obtain 
\begin{equation}
 \dot{\theta}= \frac{1}{2i} (\frac{\dot{z}}{z}-\frac{\dot{\bar{z}}}{\bar{z}}) = \frac{L}{\tilde{m}r^2} \label{theta}
\end{equation}
where we have used eqn. (\ref{ang}).
Using eqns. (\ref{r}), (\ref{z}), and (\ref{theta}), we re-express the time derivative of z as
\begin{equation}
 \dot{z} =  (\dot{r}+i\frac{L}{\tilde{m}r})\frac{z}{r} \label{z1}
\end{equation}
and hence
\begin{equation}
\overset{..}{z} = (\frac{\overset{..}{z}}{r}-\frac{L^2}{r^4}) z \label{1}  
\end{equation}
By using (\ref{eom1}) along with (\ref{1}), we re-write the equation of motion as
\begin{equation}
  \tilde{m} \overset{..}{r} = \frac{L^2}{\tilde{m}r^3} - U^{'}(r,a)
\end{equation}
Multiplying the above equation with $\dot{r}$ and integrating, we obtain
\begin{equation}
 \tilde{m}\dot{r}^2 = 2\left(E-\frac{L^2}{2\tilde{m}r^2}-U(r,a)\right) 
\end{equation}
where E is an integration constant which will be identified with total energy, later. 
The above equation can be expressed as 
\begin{equation}
  (\dot{r}(t))^2 = \frac{L^2}{\tilde{m}^2 r^2}f(r(t),a) \label{f}
\end{equation}
where $f(r,a)$ is a generic function of $r$ and the deformation parameter $a$. Taking modulus of eqn.(\ref{z1}), we
find
\begin{eqnarray}
\lvert \dot{z} \rvert^2 &=& \dot{r}^2+\frac{L^2}{\tilde{m}^2r^2}= \frac{2}{\tilde{m}} (E-U(r,a)).
\end{eqnarray}
This clearly shows that the identification of $E$ with the total energy is valid. Eqn.(\ref{f}) shows that 
$\dot{r}(t)=0$ have two possible solutions given by
\begin{equation}
  \dot{r}(t) = \pm \frac{L}{\tilde{m}r} \sqrt{f(r(t),a)}.\label{dotr}
\end{equation} 
By convention, we take the solution for ${\dot r}$ with $`+'$ sign as the solution where $r$ changes from $r_{min}$ to $r_{max}$ (call it $r_1$) and $`-'$ sign as the solution where $r$ changes from $r_{max}$ to $r_{min}$ ($r_2$).
 
Motion from $r_{min}$ to $r_{max}$ (or vice versa) is termed as a phase of the motion. Assuming that the motion starts at $r=r_{min}$ when $t=0$, we say that during the motion from $r_{min}$ to $r_{max}$ it will be in the first phase of motion, the second phase of motion will be from $r_{max}$ to $r_{min}$ and so on. Thus, one index each phase with positive integers which will make it easier for referring the phase of motion under 
consideration. This convenience is exploited by indexing each phase of motion by an integer $k \in \mathbb{Z}$ with odd $k$ corresponding to $r_1$ and even $k$ corresponding to $r_2$, respectively. Using this, we  re-express Eqn.(\ref{dotr}) compactly as
 \begin{equation}
\dot{r(t)} = (-1)^{k+1} \frac{L}{\tilde{m}r} \sqrt{f(r(t),a)}. \label{r1}.
\end{equation}
Using Eqn.(\ref{theta}) and Eqn.(\ref{dotr}), we find 
\begin{equation}
\frac{d\theta}{dr} = (-1)^{k+1} \frac{1}{r\sqrt{f(r,a)}},
\end{equation}
whose solution is
\begin{equation}
\theta(r,a) = \theta(r_0,a)+ (-1)^{k+1} \int_{r_0}^{r} \frac{1}{\rho \sqrt{f(\rho ,a)}} d\rho.    
\end{equation}
Choosing $r_0 = r_{min}$ and $\theta(r_0)=0$, we find
\begin{equation}
 \theta_{k}(r,a) = 2n \int_{r_{min}}^{r_{max}} \frac{d\rho}{\rho \sqrt{f(\rho ,a)}} + (-)^{k+1} \int_{r_{min}}^{r(t)} \frac{d\rho}{\rho \sqrt{f(\rho ,a)}},  
\end{equation} 
where $n$ is the index of phases. 
The above equation can be re-expressed as
\begin{equation}
 \theta_{k}(r,a) = 2n \Phi + (-1)^{k+1} g(r(t), a),
\end{equation}
with $k$ referring to the phase of the motion and $\Phi$ given by
\begin{equation}
 \Phi = \int_{r_{min}}^{r_{max}} \frac{d\rho}{\rho \sqrt{f(\rho, a)}}, 
\end{equation}
and
 \begin{equation}
 g(r(t),a) = \int_{r_{min}}^{r(t)} \frac{d\rho}{\rho \sqrt{f(\rho, a)}}  \label{g}.
\end{equation}
Using Eqn. (\ref{z1}) and Eqn.(\ref{dotr}), the instantaneous velocity is given as
\begin{equation}
  \dot{z} =  \left[(-1)^{k+1}\sqrt{f(r,a)}+i\right]\frac{L}{\tilde{m}r^2}z(t). \label{z2}
\end{equation}
Since under complexification $\dot{\vec{r}} \times \vec{L} \rightarrow iL\dot{z}$ we find,
\begin{equation}
 \dot{\vec{r}} \times \vec{L} \rightarrow iL\dot{z} = - \left(1+(-1)^{k}i\sqrt{f(r,a)}\right)\frac{L}{\tilde{m}r^2}z(t).  \label{laplace}
\end{equation}
This form is used in the next section to complexify the Fradkin-Bacry-Ruegg-Souriau vector.

\section{Fradkin-Bacry-Ruegg-Souriau vector in the $\kappa$-space-time}
In this section, we  generalize the approach of \cite{peres} to the $\kappa$-deformed space-time. In this approach, for arbitrary central potential, one  starts with a generic vector, which reduces to the Laplace-Runge-Lenz vector for the Kepler problem. Demanding the constancy of this vector imposes conditions on the arbitrary coefficients appearing in the definition of this vector. These conditions are expressed as coupled differential equations and by analysing these differential equations, a piecewise constant vector has been constructed explicitly. Using the results obtained in the previous section, we now generalize this construction to the $\kappa$-deformed space-time.

Our starting point in constructing the Fradkin-Bacry-Ruegg-Souriau vector using Peres approach in the $\kappa$-space-time is from  the postulate
\begin{equation}
  \vec{A} = \tilde{m}^2 \dot{\vec{r}} \times \vec{r}~~\left(\frac{r^2}{L^2} b(r,a)\right) + c(r,a) \vec{r},
\end{equation}
where $b(r,a)$ and $c(r,a)$ are generic functions of $r$ which also depend on the deformation parameter $a$ and in the limit $a\to 0$, reduces to $b(r)$ and $c(r)$. It would be safe to assume that the vector $\vec{A}$ have the same form in non-commutative setting when expressed in terms of commutative variables as all the `$a$' dependent corrections are absorbed into the definitions of ${\tilde m}$,  $b(r,a)$ and $c(r,a)$.  Note that the angular momentum appearing in the above is the $\kappa$-deformed one defined in Eqn.(\ref{ang}). After complexification, the above vector becomes 
  \begin{equation}
    \mathcal{A} = \left(\frac{r^2}{L^2} b(r,a)\right)~~ i\frac{L}{\tilde{m}}\dot{z}+c(r,a)z.
  \end{equation}
Using (\ref{z2}), we re-write the ${\mathcal {A}}$ (for the $k^{th}$ phase) as
  \begin{equation}
   \mathcal{A}_{k} = \left[c(r)-b(r)~~+~~ (-1)^{k+1} ~~i \sqrt{f(r,a)}~b(r,a)\right]~z.   
  \end{equation}
The requirement that the above vector is a constant of motion (i.e, $\dot{\mathcal{A}_{k}}=0$) results in
two coupled differential equations on c(r, a) and b(r, a). They are 
   \begin{eqnarray}
   c^{'}(r,a) -  b^{'}(r,a) + \frac{2c(r,a)-b(r,a)}{r} = 0, \label{eqn1} \\
   rf(r)c^{'}(r,a)+c(r,a)\left(\frac{rf^{'}(r,a)}{r}+f(r,a)-1\right)+b(r,a)=0, \label{eqn2}
  \end{eqnarray}
  where we have used Eqn. (\ref{z2}). After rearranging eqn.(\ref{eqn1}) as
  \begin{equation}
   \left(rc(r,a)\right)^{'} = \left(rb\right)^{'}(r,a) -c(r,a),
  \end{equation}
  and using
  \begin{eqnarray}
    c(r,a) = u^{'}(r,a);~c^{'}(r,a) = u^{''}(r,a)\label{a},
  \end{eqnarray} 
 we get
  \begin{equation}
    (rb(r,a))^{'} = ru^{''}(r,a)+2u^{'}(r,a) = (ru(r,a))^{''}.
\end{equation}
From the above equation, we find $ b(r,a) = b_{0} + u^{'}(r,a) + \frac{u(r,a)}{r}$,
 where the $b_{0}$ is the integration constant, which we set to zero. Thus the above equations
 reduces to 
  \begin{equation}
        b(r,a) = u^{'}(r,a) + \frac{u(r,a)}{r}. \label{b}
  \end{equation}
  Now using Eqn.(\ref{a}) and Eqn. (\ref{b}), eqn. (\ref{eqn2}) is re-expressed as
  \begin{equation}
       rf(r,a)u^{''}(r,a)+u^{'}(r,a)\left(\frac{rf^{'}(r,a)}{r}+f(r,a)\right) + \frac{u(r,a)}{r}=0,
\end{equation}
 which leads to
\begin{equation}
      u^{''}(r,a) + u^{'}(r,a) \left(\frac{1}{r}+\frac{f^{'}(r,a)}{f(r,a)}\right) 
+ \frac{1}{r^2f(r,a)}u(r,a)=0. 
\end{equation}
The above eqn. can be re-cast as
\begin{equation}
 u^{''}(r,a) - u^{'}(r,a) \left[\frac{1}{log(r\sqrt{f(r,a)}}\right]^{'} u^{'}(r,a) \
+ \left[\frac{1}{r\sqrt{f(r,a)}}\right]^{2}u(r,a)=0.  \label{46}
  \end{equation}  
Defining $u(r,a)=v(g(r,a))$ where $g(r,a)$ is given in Eqn.(\ref{g}), we find
  \begin{eqnarray}
  u^{'}(r,a)&=&v^{(1)}(g(r,a))g^{'}(r,a) \\
  u^{''}(r,a)&=&v^{(1)}(g(r,a))g^{''}(r,a)+v^{(2)}(g(r,a))(g^{'}(r,a))^2,
  \end{eqnarray}
where $v^{1} $ and $v^{2}$ are the first and second derivatives of $v$ with respect to the argument. Using these,
 Eqn.(\ref{46}) is expressed as
  \begin{equation}
  v^{(2)}(g(r,a)) + v(g(r,a))=0,
  \end{equation}
  This lead to the solution for $u$  given by
  \begin{equation}
  u(r,a)=v(g(r,a))=A \cos(g(r,a))+B \sin(g(r,a)). 
\end{equation}
Using this, we find
\begin{eqnarray}
 b(r,a)&=&g^{'}(r,a) (B\cos(g(r,a))-A \sin(g(r,a))), \\
 c(r,a)&=&b(r,a)+ \frac{1}{r} (A\cos(g(r,a))+B \sin(g(r,a))). 
  \end{eqnarray}
  where Eqns.(\ref{a}) and (\ref{b}) are used. Thus we see that the condition $\dot{\mathcal{A}}=0$ gives us equations which can be solved analytically (for $b$ and $c$) and thus, establish the existence of such a constant vector for generic central potentials.  
  
Thus we have derived the conditions required for the conservation of Fradkin-Bacry-Ruegg-Souriau vector in the $\kappa$ deformed space-time and have showed the existence of an integral of motion for a generic central potential in kappa-deformed space-time. 

  \section{Gorringe-Leach equation in $\kappa$-space-time}
The existence of a conserved vector, FBRS vector, for a central potentials natually lead to attempts to construct similar conserved quantities for dissipative systems also. The idea that dissipative system can be mapped to a system with no drag term is well known in the commutative case \cite{grand2}. This method have been used to obtian a conserved vector for dissipative systems. In this section, we generalize the procedure of \cite{grand2} by Grandati et.al. on Gorringe-Leach equation to the kappa-deformed space-time and establishes the existence of a piecewise conserved quantity. This is a demonstration for  the existence of conserved quantities for dissipative systems in the $\kappa$-deformed space-time. We start with a Hamiltonian in the kappa-deformed space-time for a potential that would give rise to generic form of Gorringe-Leach equation. Next, by assuming the existence of a conserved quantity, we obtain the conditions for such a conserved quantity to exist. Further we will show that, as in the 
undeformed case, this vector is piece-wise conserved.
  
 We start with Hamiltonian given in eqn.(\ref{hami}). When choosing the form of the potential, we should take care of following points:
 \begin{itemize}
 \item The corresponding equation of motion should transform into the Gorringe-Leach equation under a reparameterisation given by $ds =e^{+K(z,\overline{z})}dt$, where $s$ is the new parameter, $t$ is the old parameter and $K$ is an arbitrary function.
 \item Since the map includes an exponential, our potential should contain the same the exponential term for a neat cancellation.
\end{itemize}  
 With the above points, we consider a specific class of potentials of the form 
  \begin{equation}
  \tilde{U}(\hat{r})= U(r,a) = \int r dr (1+\alpha a E_0)^2 g_{1}(r,a) e^{2K(r,a)},
  \end{equation}
    where $g_{1}(r,a)$ and $K(r,a)$ are two arbitrary functions of radial coordinate which also depend on the deformation parameter $a$. Note that these functions reduce to the correct commutative limit as we set $a$ to zero. Without loss of generality, rewrite $K$ as $K(r,a)=K\left(r(1+a\alpha E_0\right))=K_{1}(r)+K_{2}(r,a)$ and using this define $g(r,a)$ as
\begin{equation}
g(r,a)= g_{1}(r,a)(1+\alpha a E_0)^2 e^{2K_{2}(r,a)}.
\end{equation}
Now using this equation and the above Hamiltonian, we obtain the equations of motion as 
\begin{equation}
\tilde{m}\overset{..}{\vec{r}}+g(r,a)e^{2K_{1}(r)}\vec{r}=0.
\end{equation}
where $\overset{.}{\vec{r}}=\frac{d\vec{r}}{dt}$.
Since we have restricted to two-dimensional space, as in the previous section, we re-express the above equation in terms of complex coordinates as
\begin{equation}
\tilde{m}\overset{..}{z}+g(z,\overline{z},a)e^{2K_{1}(z,\overline{z})}z=0. \label{MGLZ}
\end{equation}  
Now on-wards we will drop the suffix in $K_{1}$ for simplicity and denote it as just $K$. We would like to point out that the dependence on deformation comes through the functions $g(z,\overline{z},a)$ and $\tilde{m}$, while $K_{1}(z,\overline{z})$ is independent of deformation parameter. 

We now introduce a change of parameter from $t$ to $s$ such that
\begin{equation}
ds =e^{+K(z,\overline{z})}dt \label{para}
\end{equation}
It is interesting to see that the reparameterisation does not depend on the deformation.
We have a one-to-one correspondence between $t$ and $s=s(t, z, \overline{z})$. Using this reparameterization we re-write the equation of motion as
\begin{equation}
\tilde{m}z^{\prime \prime}+\tilde{m} K^{\prime}(z,\overline{z})z^{\prime}+g(z,\overline{z},a)z=0, \label{GLZ}
\end{equation}
where we denote $\frac{dz}{ds}$ as $z^{\prime}$. Here, we now restrict our attention to those $K$ which depends only on $r=(z \bar{z})^{\frac{1}{2}}$. We note here that the entire effect of $\kappa$-deformation is present in the deformed mass, $\tilde{m}$ and in the $g(z,\overline{z},a)$, where as the velocity dependent term is independent of the deformation and thus remains the same as in commutative space. 
One readily sees that the above equation, when expressed in terms of Cartesian coordinate system would be the generalized Gorringe-Leach equations, given by 
\begin{equation}
\tilde{m}\vec{r}^{~~\prime \prime}+ \tilde{m} h\vec{r}^{~~\prime}+g\vec{r}=0, \label{CGLZ}
\end{equation}
where h and g are arbitrary scalar function. Such an equation takes into account the drag effect on the dynamics of systems in presence of a central potential and it has been shown that they possess conserved quantities, as in the case of central potential systems in ordinary space-time. Comparing eqns. (\ref{GLZ}) and (\ref{CGLZ}), we we find that $h$ is a total derivative,i.e,
\begin{equation} 
h = K^{\prime}.
\end{equation}
This is crucial when apply inverse map to Gorringe-Leach equation to obtain an equation of the form given by eqn.(\ref{MGLZ}). If there is no such total derivative in the expression, we would not be able to achieve this mapping and this can be easily observed from the reparameterisation given in eqn.(\ref{para}).

Thus we see that the eqn.(\ref{MGLZ})is mapped to the Gorringe-Leach equation in the $\kappa$ space-time, given by eqn.\ref{GLZ}). Now it is easy to see that the angular momentum and energy are the integrals of motion for the system describing equation (\ref{MGLZ}).

The integrals of motion, angular momentum and energy, can be expressed as
\begin{equation}
\mathcal{L}= \frac{\tilde{m}}{2i} (\dot{z}\overline{z}-z\overline{\dot{z}}),
\label{LZ}
\end{equation}
and
\begin{equation}
\mathcal{E}=\frac{\tilde{m}}{2}\left\vert \dot{z}\right\vert^{2}+U(r,a)), \label{EZ}
\end{equation}   
respectively. Note that all the effects of non-commutativity is included in our equations through the factors depending on the parameter `$a$' and the structure of equations are similar to the usual commutative case apart from the modified coefficients and mass. 

As in the previous section, one introduce a piece-wise connected vector given by 
\begin{equation}
   \mathcal{A}_{k} = (c(r,a)-b(r,a)~~+~~ (-1)^{k+1} ~~i \sqrt{f(r,a)}~b(r,a))~z,  
  \end{equation}
 and it is straightforward to see that all the previous calculations remain valid for this case also. Thus we conclude that a central potential system with drag term in $\kappa$-deformed also possess a piecewise connected integral of motion as in the case of commuting space-time. 
\section{Bohlin-Arnold-Vassiliev duality and $\kappa$-deformed generalized Gorringe-Leach equations}
\begin{theorem}
Suppose the motion of a point in the complex plane is given by $w(t)$ which satisfies
\begin{equation}
\ddot{w} = - c w |w|^{\mu -1},  \qquad {\dot\omega} = \frac{d\omega}{dt}.
\end{equation} 
Then orbit of this equation can be mapped to an orbit undergoing 
\begin{equation}\label{zeqn}
z^{\prime \prime} = - {\tilde c} z |z|^{-\frac{4(\mu + 2)}{\mu + 3}} 
\qquad \hbox{ where } \qquad z^\prime = \frac{dz}{d\tau},
\end{equation}
under the transformation $z = w^{\nu}$ and Euler-Sundman's reparameterization of the type
$d\tau = |w(t)|^{2(\nu - 1)} dt$.
\end{theorem}
\noindent
{\bf Proof} Let us start from (\ref{zeqn}) 
$$
\frac{d^2z}{d\tau^2} = \frac{1}{|w|^{2(\nu - 1)}}\frac{d}{dt}\left(\frac{1}{|w|^{2(\nu - 1)}} 
\frac{dw^{\nu}}{dt} \right)
$$
$$ = \frac{\nu}{|w|^{2(\nu - 1)}} \frac{1}{{\bar w}^{\nu - 1}}\frac{d^2w}{dt^2} - 
\frac{2\nu(\nu -1)}{|w|^{2(\nu - 1)}}\frac{1}{{\bar w}^{\nu}} |\dot{w}|^2
$$
$$
= - \big(\frac{\nu c |w|^{1 + \mu}}{{\bar w}^{\nu}|w|^{2(\nu - 1)}} + 
\frac{2\nu(\nu -1)}{{\bar w}^{\nu} |w|^{2(\nu - 1)}} |\dot{w}|^2 \big)
$$
$$
= - \frac{2\nu(\nu -1)}{\bar{w}^{\nu}|w|^{2(2\nu - 1)}} 
\left( |\dot{w}|^2 + \frac{c}{2(\nu - 1)} |w|^{1 + \mu} \right).
$$
From the second term of the last expression it is clear if $( |\dot{w}|^2 + \frac{c}{2(\nu - 1)} |w|^{1 + \mu})$ has to be the energy function of $w$ equation then $ 2(\nu - 1) = 1 + \mu$ or
$ \nu = (\mu + 3)/2$.
It is clear $|z| = |w|^{\nu}$ and $ z = w^{\nu}$. If we express the denominator of the last expression
$|w|^{2(2\nu - 1)} = |z|^{\nu(1 - a)}$, we obtain $2(2\nu - 1) = \nu( 1 - a)$ or $\nu(3 + a ) = 2$.
If we substitute $ \nu = (\mu + 3)/2$ we obtain $(\mu + 3)(a + 3) = 4$, which yields
$$ (\mu + 3)( a + 3) = 4, \qquad \hbox{ and } \qquad a -1 = -\frac{4(\mu + 2)}{\mu + 3}. $$


\bigskip

If we consider the potential as a power law potential $V(w,{\bar w}) = c |w|^{1 + \mu}$, then the associated
generalized Gorringe-Leach equation is given by
\begin{equation}
\ddot{w} + \dot{K}(w,{\bar w})\dot{w} + c |w|^{\mu -1} exp(-2K(w,{\bar w})) w(t) = 0 \label{GLeqn}
\end{equation}
Thus by conformal transformation of coordinates $ z = w^{\nu}$ and Euler-Sundman reparametrization
$ d\tau = |w(t)|^{2(\nu -1)} dt$ we obtain the following equation
\begin{equation}
z^{\prime \prime} + \dot{h}(z,{\bar z})\dot{z} + {\tilde c} |z|^{a - 1} exp(-2h(z,{\bar z}) z = 0.
\end{equation}
where $\tilde{c}$ is an arbitrary constant which gives the coupling strength of the correspondent potential.

We note that the above calculations remain true for the case where the functions such as $K$,  have a dependence on the deformation parameter $`a'$. In order to see this, note that as we use expressions in the $\kappa$-deformed space, written in terms of commuting coordinates, the constant $c$ and exponential term in eqn.(\ref{GLeqn}) get $`a'$ dependence but the general structure of the equation is not modified. Thus, the results of commutative space are perfectly valid for the case of our interest i.e. in the $\kappa$-deformed space.
\subsection{Duality and $\kappa$-deformed generalized Gorringe-Leach equation }
In this subsection, we use the above result to establish the duality between the Kepler type potentials and harmonic oscillator type potentials. We first consider Gorringe-Leach equation with power potential in the kappa-deformed case and this is mapped to a Gorringe-Leach equation with a different power potential.

As in the previous section, take a power potential of the form $V(w,{\bar w}) = c |w|^{1 + \mu}(1+a\alpha E_0)^{1 + \mu}$.
Associated generalized Gorringe-Leach equation is
\begin{equation}
\ddot{w} + \dot{K}(w,{\bar w},a)\dot{w} + c |w|^{\mu -1}exp(-2K(w,{\bar w},a))  w(t) = 0,
\end{equation}
  where $c = c' (1+a\alpha E_0)^{\mu -1}$ with $c'$ being an arbitrary constant.
In the special case, where $exp(-2K(w,{\bar w},a))=exp(-2K(|w|,a))=exp(-2K(r(1+a \alpha E_0)))$, we seperate the $a$ dependence as 
\begin{eqnarray}
exp(-2K(|w|,a)) &=& exp(-2K_0(|w|))+exp(-2K_1(|w|,a)), \\
g(r,a) &=& c exp(-2K_1(|w|,a)).
\end{eqnarray}
where $K_0$ and $K_1$ are arbitrary functions of $|w|$ and $(|w|,a)$ respectively.

Using this, for the above special case, $\kappa$-deformed generalized Gorringe-Leach equation can be re-written as
\begin{equation}
\ddot{w} + \left(\dot{K_0}(|w|)+\dot{K_1}(|w|,a)\right)\dot{w} + g(r,a) |w|^{\mu -1}exp(-2K_0(|w|))  w(t) = 0.                         \label{gp}
\end{equation}
In above expression, it is easy to see that the $\kappa$-deformed equation has additional terms with specific $a$ dependence. Further, note that in the limit $a \rightarrow 0 $, we reproduce the undeformed (generalized) Gorringe-Leach equation.

We have two classes of potentials, namely the class of Harmonic-type potentials ($\mathcal{H}$ class) and class of Kepler-type potentials ($\mathcal{K}$).

$\mathcal{H}$ class is given by
\begin{equation}
g_{\mathcal{H}}(r,a) = C_{\mathcal{H}} exp(-2K_{1}(r))
\end{equation} 
and $\mathcal{K}$ class is given by
\begin{equation}
g_{\mathcal{K}}(r,a) = \frac{C_{\mathcal{K}}}{r^3} exp(-2K_{1}(r))
\end{equation} 
where $C_{\mathcal{H}}$ and $C_{\mathcal{K}}$ have the form
\begin{eqnarray}
C_{\mathcal{H}} = c (1+a\alpha E_0)^2, \\
C_{\mathcal{K}} = \frac{c}{(1+a\alpha E_0)}.
\end{eqnarray}

The labels $\mathcal{H}$ and $\mathcal{K}$ denote that they are the generic versions of harmonic potential and Kepler potential, respectively. Having defined the generic classes for the harmonic potential and Kepler potential, we are all set to validate the duality between these two classes. 

Using equation (\ref{gp}) and the above definitions, we have the following equations for $\mathcal{H}$ class:
\begin{eqnarray}
\ddot{w} + \dot{K}(|w|)\dot{w} + C_{\mathcal{K}}exp(-2K(|w|))  w(t) = 0,
\end{eqnarray}
and for $\mathcal{K}$ class, we have
\begin{eqnarray}
\ddot{w} + \dot{K}(|w|)\dot{w} + C_{\mathcal{K}} |w|^{-3} exp(-2K(|w|))  w(t) = 0.
\end{eqnarray}
Comparing this with the general expression (\ref{gp}), we have $\mu_{\mathcal{H}} = 1$ for $\mathcal{H}$ class and $\mu_{\mathcal{K}} = -2$ for $\mathcal{K}$ class. One can clearly verify that these values clearly satisfies the condition for duality given by
\begin{equation}
(\mu_{\mathcal{H}} + 3)( \mu_{\mathcal{K}} + 3) = 4.
\end{equation}
This underlies the duality relation between the Kepler-type and Harmonic oscillator-type Gorringe-Leach equations.
\section{Discussion and Conclusions}
In this paper, we have constructed conserved quantities associated with generic potentials in the $\kappa$-deformed space. The existence of a conserved vector is first established for central potentials. We have also shown the existence of a conserved vector for dissipative systems in the $\kappa$-deformed space. We have used the approach of mapping non-commuting coordinates to corresponding commuting coordinates and their functions. Thus we map the central force problem and dissipative system defined in terms of $\kappa$-deformed phase space variables to that of commutative phase space variables.

In section 2, we present a brief summary of the construction of $\kappa$-deformed Hamiltonian in terms of commutative variables. Using this, we then obtain the equation of motion valid in the $\kappa$-deformed space and re-cast it in the complex notation and obtain the expression for conserved angular momentum of this $\kappa$-deformed system. We then analyze the radial motion of this system using complex coordinate system. We have shown that the radial motion can be classified into different phases, as in commutative case. We have obtained the instantaneous velocity in terms of complex coordinate and in the limit $a$ going to zero, we get back the commutative result. These results are used in the next section. In section 3, we have constructed FBRS vector for the generic $\kappa$-deformed central potentials. We start with the most general vector which reduces to correct commutative result. This vector, for the Kepler potential is exactly same as the Laplace-Runge-Lenz vector. Using the results of the 
previous section, the requirement that this vector is a conserved quantity, is expressed as a set of coupled differential equation. We have obtained the solution to these equations, thereby explicitly constructing the FBRS vector for the general central potential in $\kappa$-deformed space. In the next section, we investigated the dynamical symmetry of system with drag term in the equation of motion in $\kappa$-deformed space. We derived the equation of motion which under reparameterization lead to a generalisation of the Gorringe-Leach equation in the $\kappa$-space-time. We have shown that even in the presence of drag term, a generalized version of Fradkin-Barcy-Ruegg-Souriau vector is an integral of motion in the $\kappa$ space-time . It is clear from the analysis presented here that the $\kappa$-deformed system will have all the dynamical symmetries of its commutative counter part. 
Further, we have established the Bohlin-Arnold-Vassiliev duality for the Gorringe-Leach equations with power potentials. This result show that Gorringe-Leach equations include classes of potentials which are generalised versions of Kepler potential and isotropic oscillator potentials in the $\kappa$-deformed space also.

\section*{Acknowledgment}
We are immensely grateful to Peter Leach for his interest and valuable comments. EH thanks S N Bose National Centre for Basic Sciences, Kolkata, India, for a visit during which a part of this work was completed. ZNS acknowledges the support from CSIR, India under the SRF scheme. 

   \end{document}